\documentclass[sigconf, nonacm, pdfa]{acmart}

\usepackage[a-2b]{pdfx}

\newcommand{\eat}[1]{}

\usepackage{latexsym}
\usepackage{amsfonts}
\usepackage{amsmath}
\usepackage{xcolor}
\usepackage{colortbl}
\usepackage{epsfig}
\usepackage{xspace}
\usepackage{graphicx}
\usepackage{subfigure}
\usepackage{paralist}
\usepackage{enumerate}
\usepackage[color,matrix,arrow,all]{xy}
\usepackage{comment}
\usepackage{booktabs}
\usepackage{balance}
\usepackage{stmaryrd}
\usepackage{pifont}
\usepackage{hhline}
\usepackage{listings}
\usepackage{array}
\usepackage{float}
\usepackage[flushleft]{threeparttable}

\usepackage{mathrsfs}
\usepackage{makecell}
\usepackage{xparse}
\usepackage{wrapfig}

\usepackage{marvosym}

\usepackage{epsfig}
\usepackage{multirow}
\usepackage{url}

\usepackage{multirow}
\usepackage{natbib}
\usepackage{graphicx}

\usepackage{listings}
\usepackage{framed}
\usepackage{xcolor}
\usepackage{color}
\usepackage{geometry}
\usepackage{changepage}
\colorlet{shadecolor}{gray!20}

\definecolor{shadecolor}{RGB}{220,220,220}

\definecolor{inputcolor}{RGB}{255,139,35}
\definecolor{outputcolor}{RGB}{120,212,252}
\definecolor{embedcolor}{RGB}{254,127,156}
\definecolor{maskcolor}{RGB}{122,128,255}
\definecolor{ecolor}{RGB}{58,149,54}

\definecolor{highcolor}{RGB}{255,153,153}
\definecolor{midcolor}{RGB}{255,204,204}
\definecolor{lowcolor}{RGB}{204,229,255}

\usepackage{tikz}
\usetikzlibrary{shapes,snakes}
\usetikzlibrary{calc}

\usepackage{algorithm} 
\usepackage{algorithmic}  
\usepackage[algo2e]{algorithm2e} 

\usepackage[export]{adjustbox}

\definecolor{green}{RGB}{0,128,0}

\definecolor{yellow}{RGB}{255,200,18}

\newcommand{\att}[1]{\textbf{\texttt{#1}}}

\newcommand{\bi}{\begin{itemize}}
\newcommand{\ei}{\end{itemize}}

\newcommand{\be}{\begin{enumerate}}
\newcommand{\ee}{\end{enumerate}}
\newcommand{\beqn}{\begin{eqnarray*}}
\newcommand{\eeqn}{\end{eqnarray*}}

\newcommand{\stitle}[1]{\vspace*{1.2pt}\noindent{\bf #1}}

\newcommand{\ie}{{i.e.,}\xspace}
\newcommand{\eg}{{e.g.,}\xspace}

\makeatletter
    \newcommand\figcaption{\def\@captype{figure}\caption}
    \newcommand\tabcaption{\def\@captype{table}\caption}
\makeatother

\tikzstyle{mybox} = [draw=black, fill=black!5, thick,
   rectangle, rounded corners, inner sep=0pt, inner ysep=6pt]
\tikzstyle{fancytitle} =[fill=black, text=white]

\NewDocumentCommand{\nan}{ mO{} }{\textcolor{blue}{\textsuperscript{\textit{Nan}}\textsf{\textbf{\small[#1]}}}}
\NewDocumentCommand{\mourad}{ mO{} }{\textcolor{blue}{\textsuperscript{\textit{Mourad}}\textsf{\textbf{\small[#1]}}}}

\newcommand{\sys}{\att{RetClean}\xspace}
\newcommand{\chat}{{GPT3.5}\xspace}

\usepackage{listings}
\usepackage{xcolor}

\definecolor{codegreen}{rgb}{0,0.6,0}
\definecolor{codegray}{rgb}{0.5,0.5,0.5}
\definecolor{codepurple}{rgb}{0.58,0,0.82}
\definecolor{backcolour}{rgb}{0.95,0.95,0.92}

\lstdefinestyle{mystyle}{
    backgroundcolor=\color{backcolour},   
    commentstyle=\color{codegreen},
    keywordstyle=\color{magenta},
    numberstyle=\tiny\color{codegray},
    stringstyle=\color{codepurple},
    basicstyle=\ttfamily\footnotesize,
    breakatwhitespace=false,         
    breaklines=true,                 
    captionpos=b,                    
    keepspaces=true,                 
    numbers=left,                    
    numbersep=5pt,                  
    showspaces=false,                
    showstringspaces=false,
    showtabs=false,                  
    tabsize=2
}

\newcommand\vldbdoi{10.14778/3685800.3685890}
\newcommand\vldbpages{4421-4424}
\newcommand\vldbvolume{17}
\newcommand\vldbissue{12}
\newcommand\vldbyear{2024}
\newcommand\vldbauthors{\authors}
\newcommand\vldbtitle{\shorttitle} 
\newcommand\vldbavailabilityurl{https://github.com/qcri/RetClean}
\newcommand\vldbpagestyle{empty} 

\begin{document}

\title{\sys: Retrieval-Based Data Cleaning Using LLMs and Data Lakes}

 \author{Zan Ahmad Naeem}
 \email{znaeem@hbku.edu.qa}
 \affiliation{%
   \institution{QCRI, HBKU}
   \country{Doha, Qatar}
 }

 \author{\mbox{Mohammad Shahmeer Ahmad}}
 \email{mohammadshahmeerah@hbku.edu.qa}
 \affiliation{%
   \institution{QCRI, HBKU}
   \country{Doha, Qatar}
 }

 \author{Mohamed Eltabakh}
 \email{meltabakh@hbku.edu.qa}
 \affiliation{%
   \institution{QCRI, HBKU}
   \country{Doha, Qatar}
 }

 \author{Mourad Ouzzani}
 \email{mouzzani@hbku.edu.qa}
 \affiliation{%
   \institution{QCRI, HBKU}
   \country{Doha, Qatar}
 }

 \author{Nan Tang}
 \email{nantang@hkust-gz.edu.cn}
 \affiliation{%
   \institution{HKUST(GZ)/HKUST}
   \country{China}
 }


\begin{abstract}
Large language models (LLMs) have shown great potential 
in data cleaning, which is a fundamental task in all modern applications.  
In this demo proposal, we demonstrate that indeed LLMs can assist in data cleaning, e.g., filling in missing values in a data table, through different approaches. 
For example, cloud-based non-private LLMs, e.g., OpenAI GPT family or Google Gemini, 
can assist in cleaning non-private datasets that encompass 
world-knowledge information (Scenario 1). 
However, such LLMs may struggle with datasets that they have never encountered before, e.g., local enterprise data, 
or when the user requires an explanation of the source of the suggested clean values. 
In that case, retrieval-based methods using RAG (Retrieval Augmented Generation) 
that complements the LLM power with 
a user-provided data source, e.g., a data lake, are a must. 
The data lake is indexed, and each time a new request comes, we  retrieve the top-$k$ relevant tuples to the user's query tuple to be cleaned and leverage LLM inference power to infer the correct  value (Scenario 2). 
Nevertheless, even in Scenario 2, sharing enterprise data with public LLMs (an externally hosted model) might not be feasible for privacy reasons. In this scenario, we showcase the practicality of  
locally hosted small LLMs in the cleaning process, especially after fine-tuning them on a small number of examples (Scenario 3). Our proposed system, \sys, seamlessly supports all three scenarios and provides a user-friendly GUI that enables the VLDB audience to explore and experiment with different LLMs and investigate their trade-offs.
\end{abstract}

\maketitle

\pagestyle{\vldbpagestyle}
\begingroup\small\noindent\raggedright\textbf{PVLDB Reference Format:}\\
\vldbauthors. \vldbtitle. PVLDB, \vldbvolume(\vldbissue): \vldbpages, \vldbyear.\\
\href{https://doi.org/\vldbdoi}{doi:\vldbdoi}
\endgroup
\begingroup
\renewcommand\thefootnote{}\footnote{\noindent
This work is licensed under the Creative Commons BY-NC-ND 4.0 International License. Visit \url{https://creativecommons.org/licenses/by-nc-nd/4.0/} to view a copy of this license. For any use beyond those covered by this license, obtain permission by emailing \href{mailto:info@vldb.org}{info@vldb.org}. Copyright is held by the owner/author(s). Publication rights licensed to the VLDB Endowment. \\
\raggedright Proceedings of the VLDB Endowment, Vol. \vldbvolume, No. \vldbissue\ %
ISSN 2150-8097. \\
\href{https://doi.org/\vldbdoi}{doi:\vldbdoi} \\
}\addtocounter{footnote}{-1}\endgroup

\ifdefempty{\vldbavailabilityurl}{}{
\vspace{.3cm}
\begingroup\small\noindent\raggedright\textbf{PVLDB Artifact Availability:}\\
The source code, data, and/or other artifacts have been made available at \url{\vldbavailabilityurl}.
\endgroup
}

\vspace{-1.8ex}
\section{Introduction}
\label{sec:intro}

Data plays a crucial role in any decision-making process, but real-world data is often riddled with missing values and errors.
Despite decades of efforts, human intervention remains the norm for practical data cleaning solutions. 
One reason is that data cleaning frequently requires domain knowledge and/or external information that may not be readily available within the data to be cleaned.

\begin{figure*}[t]
    \centering 
    \includegraphics[width=.8\textwidth, height = 6cm]{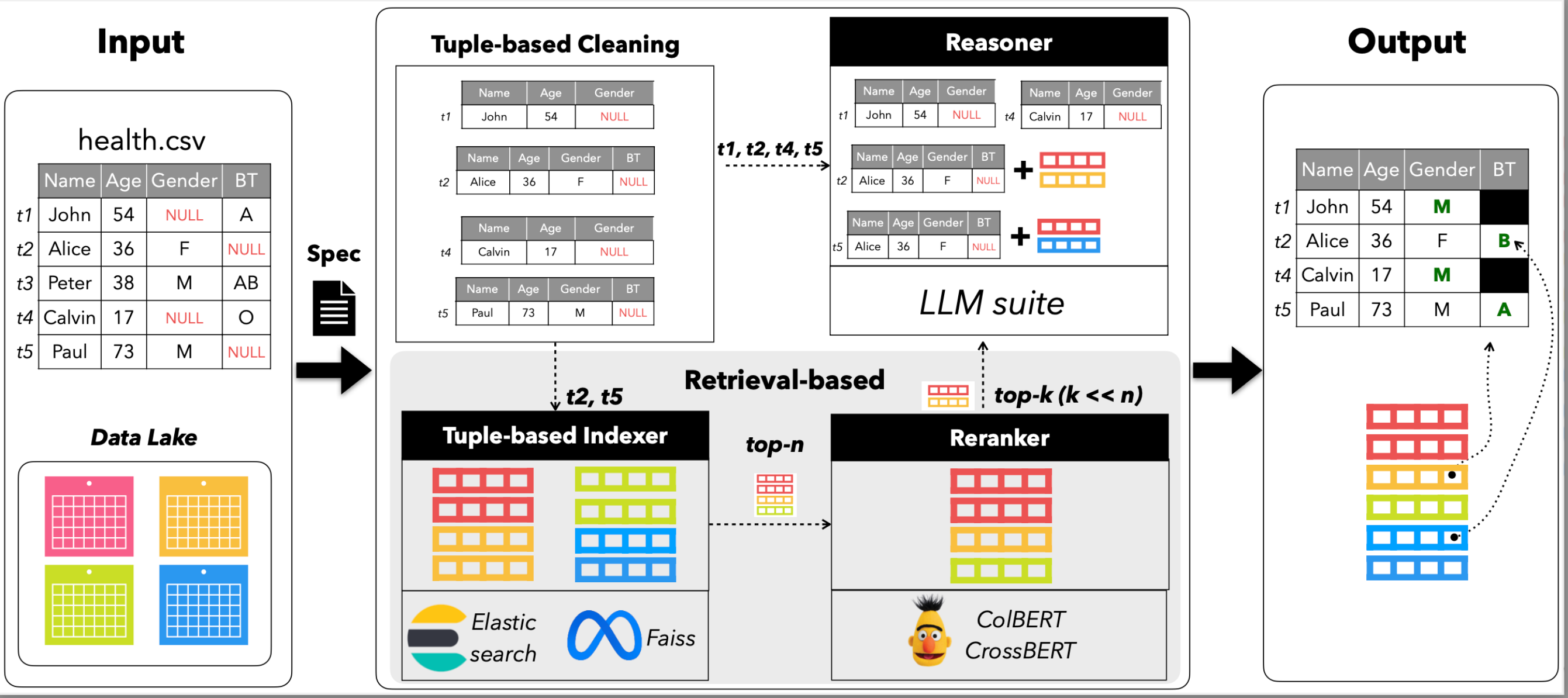}
    \caption{An Overview of \sys.}
    \label{fig:overview} 
 \end{figure*}

Large language models, e.g., OpenAI's GPT family, LLaMA, CLaude, Google Gemini, 
among others, have exhibited considerable potential in a wide range of tasks, including language-based tasks, code generation/debugging, and many others.
Recently, LLMs have been envisioned to support data wrangling~\cite{DBLP:journals/pvldb/NarayanCOR22} (\eg data cleaning and entity matching) and querying data lakes~\cite{symphony, Zhang2023LargeLM}
Furthermore, even smaller fine-tuned LLMs, e.g., 7B and 13B models, have shown comparable capabilities in several data cleaning tasks compared to the powerful GPT family~\cite{Zhang2023JellyfishAL}.  

In this demo proposal, we present \sys, an end-to-end LLM-powered data cleaning solution that supports three different data cleaning scenarios (see Figure~\ref{fig:overview}). 
Using \sys, the audience will have a hands-on experience 
in testing few different models on various datasets.

\vspace{1mm}
\stitle{Scenario 1: Data cleaning with public cloud-based LLMs (e.g., GPT family and Google Gemini).}
We first showcase how our system harnesses the capabilities of such massive LLMs.
The process is simple: the user uploads a table and identifies the column 
that may contain errors. Next, we serialize each tuple as a prompt 
and utilize the selected LLM to suggest the correct value(s) for the dirty cells.
Several prompt templates can be tried to get the best results from these LLMs.


\vspace{1mm}
\stitle{Scenario 2: Retrieval-based data cleaning with public cloud-based LLMs (e.g., GPT family and Google Gemini).}
To overcome the limitations where public LLMs may struggle with datasets that they have never
seen before and the lack of explainability of the results they return, we demonstrate the effectiveness of {\em retrieval-based} data cleaning  based on the RAG (Retrieval Augmented Generation) approach. Specifically, the domain expert provides a data lake that serves as a source of domain knowledge that could potentially contain data that can be used for cleaning the input table. \sys indexes this data lake either offline for massive data lakes 
or online  at cleaning time for small data lakes. 
Then, at cleaning time and given a dirty tuple, \sys retrieves the top-$k$ tuples 
from the data lake that could potentially help the cleaning task. 
Then, we utilize the selected LLM to make inferences about which value to use along with its source tuple--providing a better explainability compared to Scenario 1.

\vspace{1mm}
\stitle{Scenario 3: Retrieval-based data cleaning with local models (e.g., Dolly-v2-3B and LLaMA-7B).}
For both Scenarios 1 and 2, one legitimate concern when using 
externally hosted models is data privacy. 
However, locally deployable models, which are small-scale models that can be easily hosted by any organization in a variety of settings, including cloud-based environments or on-premise servers, would be ideal if one is worried about data privacy.  
Typically, they are around 3B to 13B parameter models.
To this end,  we demonstrate the effectiveness of such small models, especially when fine-tuned for a given domain. The local model takes a pair of tuples (i.e., a query tuple with a missing value and a retrieved tuple), and then infers the missing value when possible. 

\sys is designed to seamlessly support the three above scenarios. 
With its user-friendly GUI supporting different configurations, the VLDB audience can effortlessly experiment with the system and explore each of the scenarios in detail.

\section{System Architecture}
\label{sec:overview}

Figure~\ref{fig:overview} shows the architecture of \sys. 

\stitle{User Input.}
The user uploads a relational table and indicates which column(s) contain the missing values to be fixed. The user can optionally specify a subset of non-dirty pivot  columns as relevant
to the cleaning task, i.e., these columns functionally determine the values in the dirty column. 

Take the following configuration as an example 
(refer to the $3^{rd}$ column in ``health.csv'' table in Figure~\ref{fig:overview}):
\begin{lstlisting}[language=Python, caption=Not retrieval-based configuration]
table = "health.csv"
dirty_column = "Gender"
relevant_columns = ['Name','Age'] 
value = 'NULL'
is_local_model = False # use ChatGPT
\end{lstlisting}
\vspace{-2mm}
Here,  the user wants to impute the missing  values (indicated by value = NULL) 
in the Gender column. 
The Name and Age columns are identified as the pivot columns.
Assuming that these columns are not highly sensitive, the user asks \sys to use a public model 
(e.g., GPT or Gemini) to perform the missing value imputation task.

Another example of a configuration is 
(refer to the $4^{th}$ column in ``health.csv'' table):
\begin{lstlisting}[language=Python, caption=Retrieval-based configuration]
table = "health.csv"
dirty_column = "BT" # BT is blood type
relevant_columns = ALL 
value = 'NULL'
datalake = "/Users/hosp_tables/" # A folder of CSV files
is_local_model = True 
\end{lstlisting}
\vspace{-2mm}
Here, the user wants to impute the missing values in the Blood Type (BT) column. 
Such details are most probably not available as world knowledge, but could be available in a local data lake, \eg a hospital database. 
Therefore, the user opts for and specifies the use of a data lake.
The local mode flag is set to True, which indicates the use of the custom local LLM, possibly for privacy concerns.


\stitle{Non-retrieval based data cleaning.}
\sys employs a tuple-by-tuple cleaning approach. 
In the case the user opts for non-retrieval-based methods (\eg cleaning $t_1$ and $t_4$ in column Gender), \sys reads one tuple at a time and passes it to the selected LLM in the \att{Reasoner} module. 
In this case, retrieval-related modules (i.e., the Indexer and Reranker) are bypassed. 
The LLM, based the knowledge it learned from its training data, suggests values for imputation for $t_1$ and $t_4$, as depicted in the output of Figure~\ref{fig:overview}.

\stitle{Retrieval-based data cleaning.}
If the user opts for a retrieval-based method, \sys will index all tuples in the specified data lake. 
The \att{Tuple-Based Indexer} module supports both a syntactic and semantic index.
The syntactic index is implemented in Elasticsearch and use the default BM25 for similarity search.
The semantic search is performed using a vector database, namely Qdrant (https://github.com/qdrant/qdrant). We also use LlamaIndex (https://www.llamaindex.ai/) to easily connect these indexes to the LLM, either local or on the cloud.
Then, given a dirty tuple (\eg $t_2$ in the BT column), 
\sys first retrieves the top-$n$ relevant tuples, where $n$ is large enough (e.g., $n = 100$) for ensuring high recall. 
Then, the tuples are passed to the \att{Reranker} module. The main role of the \att{Reranker}
is to refine and reorder the retrieved tuples using more advanced methods, e.g., ColBERT~\cite{colbert} or CrossBERT~\cite{crossbert}, and finally produces the top-$k$ candidates, where $k \ll n$ (\eg $k = 5$), to ensure high precision. The \att{Reranker} and \att{Reasoner} are designed as separate modules to provide flexibility 
so we can easily use any arbitrary LLM from the supported LLM suite 
as the \att{Reasoner}.

Given a dirty tuple (\eg $t_2$) and top-$k$ retrieved tuples, the \att{Reasoner} module employs the selected LLM (either public or  local) to determine the appropriate retrieved tuple to use (matching step) and the value to extract for cleaning the dirty value (extraction). This step is performed in a pair-wise fashion, i.e., the query tuple and each individual retrieved tuple will form a candidate pair that is passed to the \att{Reasoner} module.
The module not only infers the imputed value (\eg blood type B for $t_2[BT]$), but also identifies the tuple and attribute from which the value is obtained (i.e., the lineage information), as demonstrated in the output of Figure~\ref{fig:overview} (the dotted lines in the output box).


\begin{figure*}[!t]
    \centering 
    \includegraphics[width=.8\textwidth,  height = 7cm]{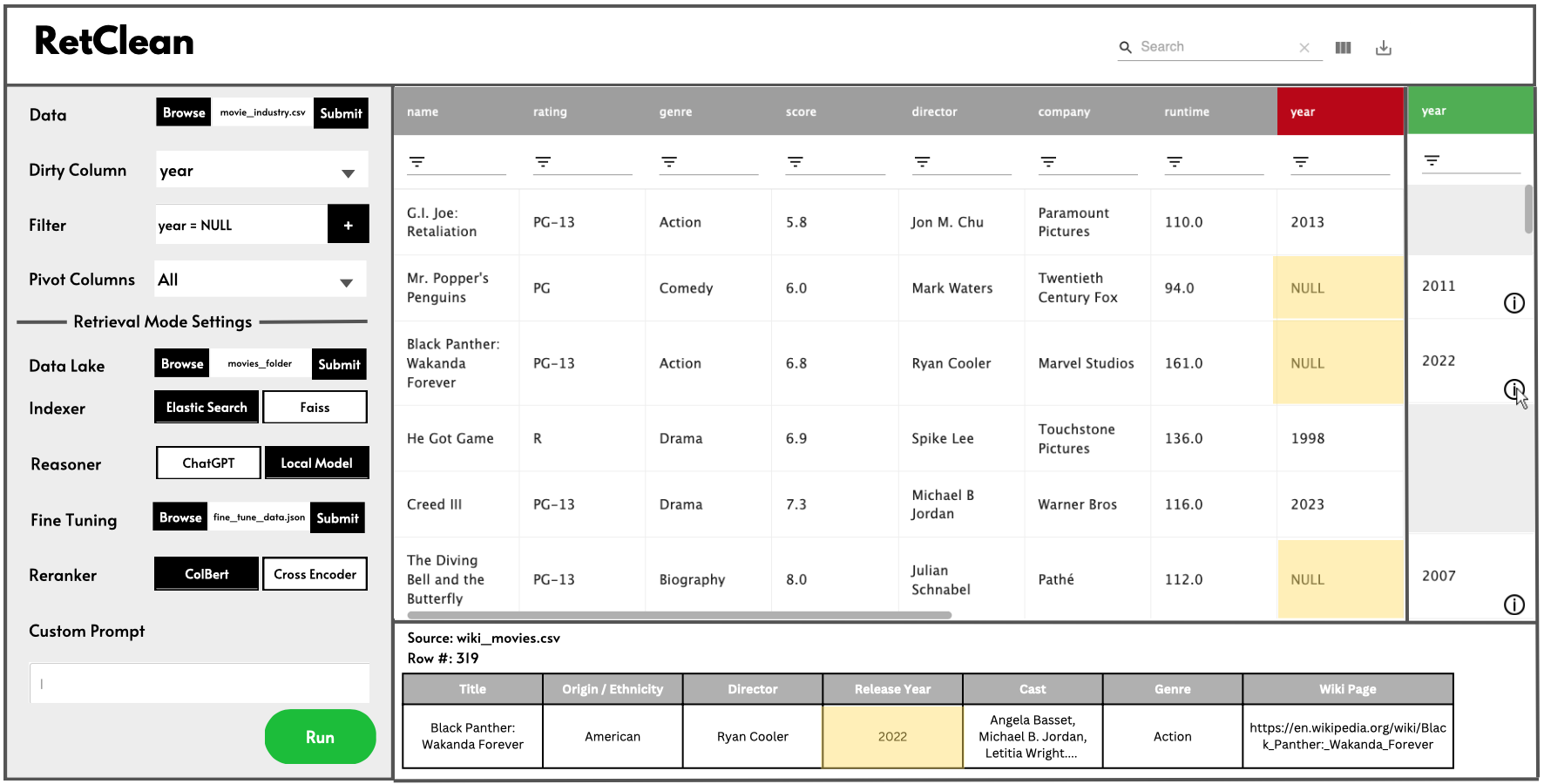}
    \vspace{-2mm}
    \caption{Demonstration Scenarios.}
    \label{fig:scenarios} 
    \vspace{-1mm}
 \end{figure*}

\section{Demonstration Scenarios}
\label{sec:scenarios}

For the demonstration, we have prepared 13 datasets from 9 different domains (\eg football matches, movies, and smartphone products), with 40 tables as the data lake. 

\stitle{Initial Setup.}
To use \sys for data cleaning, the user needs to upload a table and indicate the dirty column requiring cleaning, as shown in the top left of Figure~\ref{fig:scenarios}. 

\vspace{1mm}
\stitle{Scenario 1:} 
{\em Data cleaning with public models.}
By default, users can simply click the ``Run'' button located on the bottom left-hand side of Figure~\ref{fig:scenarios}. 
Optionally, the user can select certain columns as the pivot column(s) from the drop-down list. 
As we will demonstrate, such selection is important as it eliminates noisy (irrelevant) columns before passing the content to the backend of \sys. As such, the backend modules 
(\att{Indexer}, \att{Reranker}, \att{Reasoner}) focus only on the data that matters. 
This scenario exclusively relies on the selected LLM knowledge to generate the cleaned data values (the 
right-most green-marked column in Figure~\ref{fig:scenarios}). 



\vspace{1mm}
\stitle{Scenario 2:} 
{\em Retrieval-based data cleaning with public models.}
Consider the case that the user's data is not part of the world knowledge, e.g.,  
very recent movie data. 
In this case, the user can connect their local data lake (e.g., database or  CSV files), and request the \sys cleaning procedure to perform retrieval-based data cleaning (based on RAG). This approach utilizes different indexers, either syntactic or semantic, to retrieve relevant tuples from the data lake (see the ``Indexer'' option in Figure~\ref{fig:scenarios}). These tuples serve as the context that contains the relevant information, and then \sys leverages the LLM powerful language reasoning to extract the relevant information from these retrieved tuples.
The cleaning results are presented in the same manner as in Scenario 1. However, there is one main difference: when the user clicks the "\Info" button, the system displays the source tuple from the data lake that the value was extracted from (the bottom box in Figure~\ref{fig:scenarios}).


\vspace{1mm}
\stitle{Scenario 3:} 
{\em Retrieval-based data cleaning with local models.}
In this scenario, we use a small local model for reasoning and extracting the correct value from the retrieved tuples. Hence, the data is not sent anywhere, making it suitable for sensitive information. 

In the retrieval-based scenarios (Scenarios 2 and 3), users can optionally select a \att{Reranker} to reorder the retrieved tuples to potentially obtain higher-quality cleaning results (more details are presented in Section~\ref{sec:details}).


\section{Implementation \& Early Results}
\label{sec:details}


\subsection{Prompts to Stand-Alone Public LLM}
\sys can interface with almost any public LLMs. 
However, the default LLM  based on which the numbers presented below are obtained is \chat.
When the LLM is used  alone for data cleaning (\ie Scenario 1), each dirty tuple is converted into a prompt and sent to \chat.
By default, we use the following template: ``[attribute1 : value1 ; attribute2
: value2 ; ... attribute n : ]''. Followed by a 
question statement {\em ``what is the  value of attribute n''}.
We also allow users to provide a customized prompt template, as shown in Figure~\ref{fig:scenarios}.
Note that \chat, as a generative model, may not always give concise answers, 
we perform some post-processing on its output to extract the relevant value to be provided to the user.

\subsection{Retrieval-Based Indexer}

\sys supports two types of indexes for indexing the tuples in a user-provided data lake, namely  ElasticSearch and embeddings-based Faiss (embeddings are generated by BERT). 
Elastic search works well for retrieving tuples similar to the query tuple based on syntactic q-gram terms, while Faiss retrieves relevant tuples based on their semantic similarity. In the demonstration, we plan to use both and show their trade-offs. Indexes are typically constructed offline and available for use during the cleaning time. 




\subsection{Retrieval-Based Reranker}

The top-$n$ retrieved tuples from the index are based on a coarse-grained similarity. 
The \att{Reranker} module is designed to rerank these top-$n$ results using a more fine-grained comparison mechanism, by holistically comparing each token of the query and each token of the retrieved tuple, in order to compute a better score of the retrieved tuple. 

We adopt a ColBERT-like strategy~\cite{colbert} as the default method. To achieve this, we split the tuples into attribute-value pairs, which we treat as individual ``chunks'' for processing. Each of these chunks is independently encoded using a Sentence Transformer encoder, and a maximum similarity (\att{maxsim}) operation is performed on all chunks of the query against all chunks of the retrieved tuples. We utilize cosine similarity for the \att{maxsim} operation. This process is repeated for all retrieved tuples corresponding to each query tuple. The summed \att{maxsim} score for a retrieved tuple is used to determine its ranking score, with higher scores indicating a better match.
%





\subsection{Retrieval-Based Reasoner}

The reasoning module can be set to use either public model, e.g., GPT-4 or local model. 
In the case of the former, a prompt is created with the serialized query tuple,  serialized retrieved tuple, and a question statement corresponding to the specific data cleaning task. In this scenario, \chat may select the value from the retrieved tuple provided in the prompt, generate a value of its own, or state that no such value can be found. This process is repeated for each retrieved tuple. Thus, if we have $m$ query tuples and each gets $k$ relevant tuples from the \att{Reranker}, 
then $m*k$ prompts are sent to \chat. 
The value is then extracted from \chat's response and presented 
to the user with the source information (from the data lake). 
Each prompt sent to \chat consists of two cascaded questions. The first question is 
{\em ``Do these two tuples relate to the same exact entity?''} with potential answers 
of Yes or No. Only if the answer is Yes, the second question, within the same prompt,  
becomes active in the form of {\em ``what is the value for ...''}.

For the local model case,  
we developed two custom RoBERTa-based~\cite{roberta} models; the {\em Matcher} 
and {\em Extractor} models, both are encapsulated inside the \att{Reasoner} module. 
The {\em Matcher} model is trained to take two serialized tuples (query \& retrieved) 
and outputs a Boolean value indicating whether or not they \textit{match}.
Here, matching implies two conditions;  the two tuples are about the same entity, e.g., same movie, player, book, and the target dirty attribute is present in the retrieved tuple, even if it is not an exact syntactic match.  
Notice that the pivot columns significantly help in this task because the {\em Matcher} focuses only on the columns that matter. 
In our experiments, we also found that for unseen datasets and schemas, the performance of the {\em Matcher} model significantly improves if it is fine-tuned on a small number of examples, e.g., 10 or 20 samples from the new dataset. 
A pair of tuples that passes the  {\em Matcher} feeds the  {\em Extractor} model.   

For the {\em Extractor} model, \sys supports out-of-the-box LLMs, 
e.g., LLaMA 3 or Qwen, as well as fine-tuned LLMs
that are trained to identify and extract the desired value from the retrieved tuple.    

\subsection{Preliminary Results}
In Table~\ref{Table:earlyresults}, we present the results of our experiments on four datasets using the three cleaning approaches.
The dirty columns with missing values for each dataset are: "Country" for Cricket Players (CP), "Genetic Order" for Animals (AN), "Age Rating" for Shows-Movies (SM), 
and "Department" for QCRI Personnel (QP).   
For the retrieval-based techniques, we manually constructed a data lake of 12 tables covering the four domains. 
For the CP dataset, which is mostly part of the world knowledge, we observe that the stand-alone \chat as well as the retrieval-based techniques all perform well. 
However, for the AN and SM datasets where the missing information is harder to find, \eg genetic information for different animals, or show ratings for unpopular shows, the retrieval-based techniques are superior. 
For a very domain-specific dataset (\eg the QP dataset), it is clear that 
the stand-alone GPT model is useless, and the cleaning process has to rely on  the private data lake. 
It is worth highlighting that our developed custom model (Scenario 3)
is competitively effective in its inference power to \chat. 

\small{
\begin{table}[h!]
\caption{Accuracy Scores for RetClean Experiments.}
\vspace{-4mm}
\begin{center}
\begin{tabular}{|l|l|l|l|l|}
\hline
Dataset & \#Tuples & Scenario 1 & Scenario 2 & Scenario 3 \\ 
 \hline\hline
Cricket Players (CP)& 100 & 97\% & 96\% & 97\% \\\hline 
Animals (AN) & 100 & 79\% & 96\% & 96\%  \\\hline
Shows-Movies (SM) & 100 & 27\% & 57\% & 74\% \\\hline
QCRI Personnel (QP) & 18 & 5\% & 94\% & 88\% \\\hline
\end{tabular}
\label{Table:earlyresults}
\end{center}
\end{table}
}

\balance
\bibliographystyle{ACM-Reference-Format}
\bibliography{DA}

\end{document}